\magnification\magstephalf
\parskip3pt
\baselineskip14pt
\def\AW{Addison\kern.1em--Wesley}
\font\sc=cmcsc10 
\def\bib[#1] {\par\noindent\hangindent 40pt\hbox to20pt{[#1]\hfil}}

\def\GW{1} 
\def\JKLP{2} 
\def\Kii{3} 
\def\Kiii{4}
\def\KW{5} 
\def\Kr{6} 
\def\MR{7} 
\def\R{8} 
\def\SF{9} 
\def\VP{10} 
\def\Wr{11} 

\centerline{\bf Linear Probing and Graphs}
\bigskip
\centerline{Donald E. Knuth, Stanford University}
\bigskip
\centerline{\sc Dedicated to Philippe Patrick Michel Flajolet}
\bigskip

{

\baselineskip12pt

{\narrower\smallskip\noindent
{\bf Abstract.} Mallows and Riordan showed in 1968 that labeled trees with
a small number of inversions are  related to labeled graphs that
are connected and sparse. Wright enumerated sparse connected graphs in
1977, and Kreweras related the inversions of trees to the so-called
``parking problem'' in 1980. A~combination of these three results leads to
a surprisingly simple analysis of the behavior of hashing by linear
probing, including higher moments of the cost of successful search.
\smallskip}

}

The well-known algorithm of {\it linear probing\/} for $n$ items in $m>n$
cells can be described as follows: Begin with all cells $(0,1,\ldots,m-1)$
empty; then for $1\leq k\leq n$, insert the $k$\/th item into the first
nonempty cell in the sequence $h_k, (h_k+1)\bmod~m, (h_k+2)\bmod
m,\ldots\,$, where $h_k$ is a random integer in the range $0\leq
h_k<m$. (See, for example, 
[\Kiii, Algorithm~6.4L].)

The purpose of this note is to exhibit a surprisingly simple solution to a
problem that appears in a recent book by Sedgewick and Flajolet 
[\SF]:

{\narrower\smallskip\noindent
{\bf Exercise 8.39}\ \  Use the symbolic method to derive the EGF of the
number of probes required by linear probing in a successful search, for
fixed~$M$.
\smallskip}

\noindent
The authors admitted that they did not know how to solve the problem, in
spite of the fact that a ``symbolic method'' was the key to the analysis of
all the other algorithms in their book. Indeed, the second moment of the
distribution of successful search by linear probing was unknown 
when [\SF] was published in~1996.

If the $k$\/th item is inserted into position~$q_k$, the quantity
$d=\sum_{k=1}^n(q_k-h_k)\bmod m$ is the {\it total displacement\/} of the
items from their hash addresses. The average number of probes needed in a
successful search is then $1+d/n$. Our goal in the following is to study
the probability distribution of~$d$ as a function of the table size~$m$ and
the number of items~$n$.

\medskip\noindent
{\bf 1. Generating functions.}
Let $D_{mn}(x)=\sum x^d$, summed over all~$m^n$ possible hash sequences
$h_1\ldots h_n$, and let $F_{mn}(x)$ be the same sum restricted to hash
functions that are {\it confined}, in the sense that linear probing with
$h_1\ldots h_n$ will leave cell~0 unoccupied.

Given $h_1\ldots h_n$, the $m$~hash sequences $\bigl((h_1+j)\bmod
m \ldots (h_n+j)\bmod m\bigr)$ for $0\leq j<m$
all lead to the same total
displacement~$d$. And exactly $(m-n)/m$ of them will be confined, in the
sense above. Therefore $D_{mn}(x)={m\over m-n}\,F_{mn}(x)$, and the
probability generating function for $d$~is
$${D_{mn}(x)\over D_{mn}(1)}={F_{mn}(x)\over F_{mn}(1)}\,.\eqno(1.1)$$
The quantity $F_{mn}(x)$ is easier to deal with than $D_{mn}(x)$, since
 linear
probing does not ``wrap around'' when the hash sequence is confined. We
obviously have $0<h_k\leq q_k<m$ in a confined sequence; therefore
remainders mod~$m$ are not actually taken and the behavior is simpler.

The special case of confined linear probing in which $m=n+1$ has been
called the {\it parking problem\/}
[\KW],
because we can think of $n$~cars that try to park in $n$~consecutive
spaces, where the $k$\/th car starts its search in position~$h_k$. The
number of sequences
 $h_1\ldots h_n$ such that all cars are successfully parked is
the number of confined hash sequences, namely ${m-n\over
m}m^n=(n+1)^{n-1}$, when $m=n+1$. We will write 
$$F_n(x)=F_{n+1,n}(x)\eqno(1.2)$$
for the generating function of total displacement in the parking problem.

The general case is clearly related to the special case $m=n+1$ by
$$F_{n+r,n}(x)=\sum_{n_1+n_2+\cdots +n_r=n}\,
{n!\over n_1!\,n_2!\,\ldots\,n_r!}\,
F_{n_1}(x)\, F_{n_2}(x)\,\ldots\,F_{n_r}(x)\,,\eqno(1.3)$$
because every confined hash sequence leaves $r$~cells
$$\{0,n_1+1,n_1+n_2+2,\ldots ,n_1+\cdots +n_{r-1}+r-1\}$$
 empty, and defines
parking sequences on blocks of sizes $n_1+1,n_2+1,\ldots,n_r+1$ for some
nonnegative integers
$n_1,n_2,\ldots,n_r$. The number of ways to fit such subsequences
into $h_1\ldots h_n$ is the multinomial coefficient $n!/n_1!n_2!\ldots n_r!$. 

Let
$$F(x,z)=\sum_{n\geq 0} F_n(x)\,{z^n\over n!}\eqno(1.4)$$
generate the displacements of successfully parked cars.
Equation (1.3) tells us that
$${F_{mn}(x)\over n!}=[z^n]\,F(x,z)^{m-n}\,;\eqno(1.5)$$
hence the bivariate generating function $F(x,z)$ is the key to the
distribution of total displacement.

\medskip\noindent
{\bf 2. Solution to the parking problem.}
Suppose $h_1\ldots h_n$ is a confined hash sequence for the special case
$m=n+1$, with $n\geq 1$. This holds if and only if $h_n\geq 1$ and
$h_1\ldots h_{n-1}$
leaves cells~0 and~$k$ empty for some~$k$ in the range $h_n\leq k\leq n$.
The sequence $h_1\ldots h_{n-1}$ then decomposes into
parking subsequences for $k-1$ and $n-k$ cars.

Therefore, by arguing as in (1.3) above, we see that the polynomials 
$F_n(x)$ satisfy the recurrence
$$F_n(x)=\sum_{k=1}^n\,{n-1\choose k-1}(1+x+\cdots +x^{k-1})\,F_{k-1}(x)\,
F_{n-k}(x)\,.\eqno(2.1)$$
(The factor $1+x+\cdots +x^{k-1}$ corresponds to the displacement of the
$n$\/th car, while ${n-1\choose k-1}$ is the number of ways to mix the two
subsequences.) The first few values are
$$\eqalignno{F_0(x)&=1\,;\cr
F_1(x)&=1\,;\cr
F_2(x)&=2+x\,;\cr
F_3(x)&=6+6x+3x^2+x^3\,.&(2.2)\cr}$$

Recurrence (2.1) can be put into a more user-friendly form if we write
$$A_n(x)=(x-1)^n\,F_n(x)\,.\eqno(2.3)$$
Then $A_0(x)=1$, and for $n>0$ we have
$$A_n(x)=\sum_{k=1}^n\,{n-1\choose
k-1}(x^k-1)A_{k-1}(x)A_{n-k}(x)\,.\eqno(2.4)$$ 

For fixed $x$, this
recurrence can be analyzed by using the exponential generating functions
$$\eqalignno{A(z)&=\sum_{n=0}^{\infty}A_n(x)\,{z^n\over n!}\,,&(2.5)\cr
\noalign{\smallskip}
B(z)&=\sum_{n=1}^{\infty}B_n(x)\,{z^n\over n!}\,,&(2.6)\cr}$$
where
$$B_n(x)=(x^n-1)A_{n-1}(x)\,,\eqno(2.7)$$
because (2.4) is then equivalent to
$$A(z)=e^{B(z)}\,,\eqno(2.8)$$
by Euler's well-known formula for power series exponentiation (see, for
example, exercise 4.7--4 in~[3]).

Now (2.6) and (2.7) tell us that
$$B(z)=C(xz)-C(z)\,,\eqno(2.9)$$
where
$$\eqalignno{C(z)&=\sum_{n=1}^{\infty}C_n(x)\,{z^n\over n!}\,,&(2.10)\cr
\noalign{\smallskip}
C_n(x)&=A_{n-1}(x)\,;&(2.11)\cr}$$
and we have
$$C'(z)=\sum_{n=1}^{\infty}C_n(x)\,{z^{n-1}\over (n-1)!}=\sum_{n=0}^{\infty}
A_n(x)\,{z^n\over n!}=A(z)\,.\eqno(2.12)$$
In other words $C'(z)=e^{C(xz)-C(z)}$; and if we set
$$G(z)=e^{C(z)}\eqno(2.13)$$
we find
$$G'(z)=C'(z)G(z)=e^{C(xz)}=G(xz)\,.\eqno(2.14)$$
But this functional relation is easy to solve, for if we set
$$G(z)=\sum G_n(x)\,{z^n\over n!}\eqno(2.15)$$
the relation $G(xz)=G'(z)$ says simply that $x^nG_n(x)=G_{n+1}(x)$. 
Therefore
$$G(z)=\sum_{n=0}^{\infty}x^{n(n-1)/2}\,{z^n\over n!}\,,\eqno(2.16)$$
and we have deduced that
$$\sum_{n=1}^{\infty}(x-1)^{n-1}\,F_{n-1}(x)\,{z^n\over n!}=C(z)=
\ln\sum_{n=0}^{\infty}x^{n(n-1)/2}\,{z^n\over n!}\,.\eqno(2.17)$$

\medskip\noindent
{\bf 3. Connected graphs.}
We are interested in the behavior of $F_n(x)$ near $x=1$, so it is
convenient to write $x=1+w$. Then (2.17) becomes
$$\sum_{n=1}^{\infty}w^{n-1}\,F_{n-1}(1+w)\,{z^n\over
n!}=\ln\sum_{n=0}^{\infty} (1+w)^{n(n-1)/2}\,{z^n\over n!}\,.\eqno(3.1)$$
Aha---the right side of this equation is well known as the exponential
generating function for labeled connected graphs
[\R]. Thus we have
$$w^{n-1}\,F_{n-1}(1+w)=C_n(1+w)=\sum w^{{\rm edges}(G)}\,,\eqno(3.2)$$
where the sum is over all connected graphs on $n$ labeled vertices.

From this interpretation of $C_n(w)$, we see that
$$F_n(1+w)=C_{n,n+1}+w\,C_{n+1,n+1}+w^2\,C_{n+2,n+1}+\cdots\;,\eqno(3.3)$$
where $C_{m,n}$ is the number of connected labeled graphs on $n$ vertices
and $m$~edges. In particular, $C_{n,n+1}$ is $(n+1)^{n-1}$, the number of
labeled trees on $n+1$ vertices; this checks with the value of $F_n(1)$
that we already knew.

\medskip\noindent
{\bf 4. Sparse connected graphs.}
Let
$$W_k(z)=\sum_{n=1}^{\infty}\,C_{n-1+k,n}\,{z^n\over n!}\eqno(4.1)$$
be the generating function for $k$-cyclic components of a labeled graph;
thus $W_0(z)$ generates unrooted trees, $W_1(z)$ generates connected
components that have exactly one cycle, $W_2(z)$ generates bicyclic
components, and in general $W_k(z)$ generates connected graphs that have
$k-1$ more edges than vertices. From (3.3) and (1.4) we have
$$F(1+w,z)=W'_0(z)+wW'_1(z)+w^2W'_2(z)+\cdots \;.\eqno(4.2)$$
E. M. Wright [\Wr] showed how to compute the $W$'s systematically, and
proved that they are all expressible in terms of the {\it tree function\/}
$$T(z)=\sum_{n=1}^{\infty}\,n^{n-1}\,{z^n\over n!}\,,\eqno(4.3)$$
which generates  rooted trees. $\bigl($See [\JKLP] 
for simplifications and
extensions of Wright's results. In that paper, $W_0(z)$, $W_1(z)$, and
$W_2(z)$ are called respectively $\hat{U}(z)$, $\hat{V}(z)$, and
$\hat{W}(z)$.$\bigr)$.

The known results about $W_k(z)$ for small $k$ show that we have
$$F(1+w,z)={T(z)\over z}\,f\bigl(w,T(z)\bigr)\,,\eqno(4.4)$$
where $f(w,t)$ has the following leading terms:
$$\eqalignno{f(w,t)=1&+w\,{t^2\over 2(1-t)^2}\cr
\noalign{\smallskip}
&\null+ w^2\left({5\over 24}\;{t^4\over (1-t)^5}\,(5-2t)+{1\over
4}\;{t^3\over (1-t)^4}\,(4-2t)\right)\cr
\noalign{\smallskip}
&\null+w^3\left({5\over 16}\;{t^7\over (1-t)^8}\,(8-2t)+{55\over 48}\;
{t^6\over (1-t)^7}\,(7-2t)\right.\cr
\noalign{\smallskip}
&\qquad\qquad\qquad\qquad\qquad
\null + {73\over 48}\;{t^5\over (1-t)^6}\,(6-2t)+{3\over
4}\;{t^4\over (1-t)^5}\, (5-2t)\cr
\noalign{\smallskip}
&\qquad\qquad\qquad\qquad\qquad
\null +\left.{1\over 24}\;{t^3\over (1-t)^4}\,(4-2t)\right)\cr
\noalign{\smallskip}
&\null +w^4\left({1105\over 1152}\;{t^{10}\over (1-t)^{11}} (11-2t)
+\cdots\,\right)\cr
\noalign{\smallskip}
&\null+\cdots&(4.5)\cr}$$
$\bigl($See formula (8.13) in [\JKLP], and use the fact that 
$zT'(z)=T(z)/\bigl(1-T(z)\bigr)$.$\bigr)$

\medskip\noindent
{\bf 5. Application to linear probing.}
We can now put everything together and calculate factorial moments of the
distribution of total displacement when $n$~items are inserted into
$m$~cells by linear probing. The tree function has a wonderful property
that leads to considerable simplification, thanks to Lagrange's inversion
formula and the identity $T(z)=ze^{T(z)}$:
$$\eqalignno{[z^n]\,&F(1+w,z)^{m-n}\cr
\noalign{\smallskip}
&\qquad =[z^n]\;{T(z)^{m-n}\,f\bigl(w,T(z)\bigr)^{m-n}\over z^{m-n}}\cr
\noalign{\smallskip}
&\qquad =[z^m]\,T(z)^{m-n}\,f\bigl(w,T(z)\bigr)^{m-n}\cr
\noalign{\smallskip}
&\qquad =[t^n]\,e^{mt}(1-t)\,f(w,t)^{m-n}\,.&(5.1)\cr}$$
(See [\Kii], third edition, exercise 4.7--16, for a simple algorithmic
proof of Lagrange's formula.)

We will need to use the functions
$$Q_r(m,n)={r\choose 0}+{r+1\choose 1}\,{n\over m}+{r+2\choose
2}\,{n(n-1)\over m^2}+\cdots= {_2F_0}(r+1,-n\,;\,;-1/m)\;,\eqno(5.2)$$
which are known to appear in the analysis of linear probing (see [\Kiii],
Theorem~K); they have the simple generating function
$$\sum_{n=0}^{\infty}\,Q_r(m,n)\,{t^n\over n!}={e^t\over (1-t/m)^{r+1}}\,.
\eqno(5.3)$$

The formulas above now allow us to compute the expected total displacement
as follows, using (5.3) and (4.5):
$$\eqalignno{&{[wz^n]\,F(1+w,z)^{m-n}\over [z^n]\,F(1,z)^{m-n}}\cr
\noalign{\smallskip}
&\qquad\qquad ={[t^n]\,e^{mt}(1-t)(m-n)\,t^2\!/\bigl(2(1-t)^2\bigr)\over
[t^n]\,e^{mt}(1-t)}\cr
\noalign{\smallskip}
&\qquad\qquad ={{1\over 2}\,(m-n)\,[t^n]\,e^{mt}t\bigl(1/(1-t)-1\bigr)\over
m^n\!/n! - m^{n-1}\!/(n-1)!}\cr
\noalign{\smallskip}
&\qquad\qquad={{1\over 2}(m-n)\,m^{n-1}\bigl(Q_0(m,n-1)-1\bigr)/(n-1)!\over
(m-n)\,m^{n-1}\!/n!}\cr
\noalign{\smallskip}
&\qquad\qquad ={n\over 2}\,\bigl(Q_0(m,n-1)-1\bigr)\,.&(5.4)\cr}$$
This agrees with the known result that a successful search requires 
${1\over 2}\bigl(Q_0(m,n-1)+1\bigr)$ probes, on the average 
[\Kiii, Theorem~K].

Moreover, a similar calculation gives
$$\eqalignno{&{[w^2z^n]\,F(1+w,z)^{m-n}\over [z^n]\,F(1,z)^{m-n}}\cr
\noalign{\smallskip}
&\qquad={n(n-1)(n-2)\over 24m^2}\,\bigl(15Q_3(m,n-3)+(4+3m-3n)Q_2(m,n-3)\cr
\noalign{\smallskip}
&\qquad\qquad\qquad\qquad\qquad\qquad\null
+(5-3m+3n)Q_1(m,n-3)\bigr)\,.&(5.5)\cr}$$
This is the expected value of ${d\choose 2}$, from which of course we
obtain the expected value of~$d^2$ by doubling and adding~(5.4). All
moments can in principle be obtained in this way, although the expressions
get more and more complicated.

Formulas such as (5.5) can be rewritten in many ways using the identities
$$\eqalignno{rQ_r(m,n)&=mQ_{r-2}(m,n)-(m-n-r)Q_{r-1}(m,n)\,;&(5.6)\cr
rQ_r(m,n)&=mQ_{r-1}(m,n+1)-mQ_{r-1}(m,n)\,;&(5.7)\cr
nQ_r(m,n-1)&=mQ_r(m,n)-mQ_{r-1}(m,n)\,.&(5.8)\cr}$$
However, none of these transformations seems to convert (5.5) into a
substantially simpler formula.

\medskip\noindent
{\bf 6. Related work.}
Germain Kreweras 
[\Kr]
discussed the polynomials $F_n(x)$ at length, showing that they are the
generating functions for ``suites majeures,'' which are equivalent to
parking sequences with displacements enumerated. He also showed that
$F_n(-1)$ is the number of ``up-down'' permutations, and that $F_n(x)$ is
the generating function for inversions in a labeled tree of $n+1$
nodes. The concept of inversions in trees was first defined by Colin
Mallows and John Riordan
[\MR],
who established their relation to connected graphs. Thus, all of the main
ideas of sections 2, 3,~4 were already in the literature, waiting to be
applied to the analysis of linear probing.

A one-to-one correspondence that maps labeled trees on $\{0,1,\ldots,n\}$
with $k$~inversions bijectively into parking sequences on $\{1,\ldots,n\}$
with $k$~displacements appears in
[\Kiii, second edition, answer to exercise 6.4--31].
A~beautiful construction that uses depth-first search to establish~(3.2), by
relating each $n$-node 
tree with $k$~inversions to $2^k$~connected graphs having
$w^n(1+w)^k$ edges, was found by Ira Gessel and Da-Lun Wang 
[\GW].
Therefore the relation between linear probing and graphs can be made quite
explicit, although there is apparently no really simple connection.

The expected value of $d^2$ was first obtained by Alfredo Viola and
Patricio Poblete
[\VP],
who discovered a formula equivalent to (5.5) about one week before the
author had independently carried out the calculations above. Their starting
point was equivalent to the symmetry-breaking strategy of section~1; their
other methods provide an interesting alternative to those of the present
note.

\medskip\noindent
{\bf 7. Personal remarks.}
The problem of linear probing is near and dear to my heart,
because I~found it immensely satisfying to deduce (5.4) when I~first
studied the problem in 1962. Linear probing was the first algorithm 
that I~was able
to analyze successfully, and the experience had a significant effect on my
future career as a computer scientist. None of the methods available in
1962 were powerful enough to deduce the expected square displacement, much
less the higher moments, so it is an even greater pleasure to be able to
derive such results today from other work that has enriched the field of
combinatorial mathematics during a period of 35~years.

It is also gratifying to know that the field of algorithmic analysis has
matured to the point where researchers in different parts of the world are
now able to resolve such difficult problems working independently.

The reader will note that Sedgewick and Flajolet's exercise 8.39 has not
truly been solved, strictly speaking, because we have not found the EGF
$\sum_{n=0}^{m-1}\,F_{mn}(x)\,z^n\!/n!$ as requested. However, Sedgewick
and Flajolet should be happy with any analysis of linear probing that uses
symbolic methods associated with generating functions in an informative
way.

I thank the referees for their perceptive remarks and valuable suggestions.

Finally, I wish to pay tribute to my secretary of more than twenty-five
years, Phyllis Astrid Benson Winkler, who is retiring this year. The
present paper is the last of more than one hundred that she has typed and
typeset beautifully for me at Stanford.

\bigskip
\centerline{References}

\smallskip
\bib
[\GW]
Ira Gessel and Da-Lun Wang, ``Depth-first search as a combinatorial
correspondence,'' {\sl Journal of Combinatorial Theory\/} (A) {\bf 26}
(1979), 308--313.

\smallskip
\bib
[\JKLP]
Svante Janson, Donald E. Knuth, Tomasz {\L}uczak, and Boris Pittel, ``The
birth of the giant component,'' {\sl Random Structures and Algorithms\/ \bf
4} (1993), 233--358.

\smallskip
\bib
[\Kii]
Donald E. Knuth, {\sl Seminumerical Algorithms}, third edition, 
(Reading, Massachusetts: \AW, 1997).

\smallskip
\bib
[\Kiii] Donald E. Knuth, {\sl Sorting and Searching}, second edition,
(Reading, Massachusetts: \AW, 1998).

\smallskip
\bib
[\KW]
Alan G. Konheim and Benjamin Weiss, ``An occupancy discipline and
applications,'' {\sl SIAM Journal on Applied Mathematics\/ \bf 14} (1966),
1266--1274.

\smallskip
\bib
[\Kr]
G. Kreweras, ``Une famille de polyn\^omes ayant plusieurs propri\'et\'es
\'enum\'eratives,'' {\sl Periodica Mathematica Hungarica\/ \bf 11}
 (1980), 309--320.

\smallskip
\bib
[\MR]
C. L. Mallows and John Riordan, ``The inversion enumerator for labelled
trees,'' {\sl Bulletin of the American Mathematical Society\/ \bf 74}
(1968), 92--94.

\smallskip
\bib
[\R]
R. J. Riddell, Jr., {\sl Contributions to the Theory of Condensation\/}
(Ann Arbor: University of Michigan, 1951). The main results of this
dissertation were published as R.~J. Riddell,~Jr., and G.~E. Uhlenbeck,
``On the theory of the virial development of the equation of the state of
monoatomic gases,'' {\sl Journal of Chemical Physics\/ \bf 21} (1953),
2056--2064. 

\smallskip
\bib
[\SF]
Robert Sedgewick and Philippe Flajolet, {\sl An Introduction to the Analysis
of Algorithms\/}  (Reading, Massachusetts: \AW, 1996).

\smallskip
\bib
[\VP]
Alfredo Viola and Patricio V. Poblete, ``Analysis of the total displacement
in linear probing hashing,'' presented at the third Dagstuhl Seminar in
Analysis of Algorithms (9~July 1997).

\smallskip
\bib
[\Wr]
E. M. Wright, ``The number of connected sparsely edged graphs,'' {\sl
Journal of Graph Theory\/ \bf 1} (1977), 317--330.

\bye